 \documentclass[page-classic]{epl2} % for one column style

\usepackage{epsfig}

\long\def\revision#1{#1}

\def\be{\begin{equation}}
\def\ee{\end{equation}}
\def\bey{\begin{eqnarray}}
\def\eey{\end{eqnarray}}

\def\tmod{|\tau|}

\def\date{\number\day\space\month\space\number\year}
\def\tmod{|\tau|}

\def\phizero{\phi(0)}
\def\psizero{\psi(0)}
\def\hzero{h(0)}
\def\phil{\phi(l)}
\def\psil{\psi(l)}

\def\phizero{\phi_0}
\def\psizero{\psi_0}
\def\hzero{h_0}
\def\phil{\phi}
\def\psil{\psi}

\def\yT{y_\phi}
\def\yH{y_h}
\def\ypsideux{y_{\psi^2}}
\def\ypsitrois{y_{\psi^3}}
\def\yTpsi{y_{\phi\psi}}
\def\yHpsi{y_{h \psi}}

\def\yTn{y_{\phi_n}}
\def\yTun{y_{\phi_1}}
\def\yTdeux{y_{\phi_2}}
\def\yTtrois{y_{\phi_3}}
\def\yHn{y_{h_n}}
\def\yHun{y_{h_1}}
\def\yHdeux{y_{h_2}}
\def\yHtrois{y_{h_3}}
\def\Eq#1{Eq.~(\ref{#1})}

\def\tmod{|\tau|}

\def\phizero{\phi(0)}
\def\psizero{\psi(0)}
\def\hzero{h(0)}
\def\phil{\phi(l)}
\def\psil{\psi(l)}

\def\phizero{\phi_0}
\def\psizero{\psi_0}
\def\hzero{h_0}
\def\phil{\phi}
\def\psil{\psi}

\def\yT{y_\phi}
\def\yH{y_h}
\def\ypsideux{y_{\psi^2}}
\def\ypsitrois{y_{\psi^3}}
\def\yTpsi{y_{\phi\psi}}
\def\yHpsi{y_{h \psi}}

\def\yTn{y_{\phi_n}}
\def\yTun{y_{\phi_1}}
\def\yTdeux{y_{\phi_2}}
\def\yTtrois{y_{\phi_3}}
\def\yHn{y_{h_n}}
\def\yHun{y_{h_1}}
\def\yHdeux{y_{h_2}}
\def\yHtrois{y_{h_3}}
\def\<{\langle}
\def\>{\rangle}
\def\Eq#1{Eq.~(\ref{#1})}

\title{A study of logarithmic corrections and universal amplitude
    ratios in the two-dimensional 4-state Potts
    model}
\shorttitle{Amplitude ratios in 4-state Potts model}
\author{L.N. Shchur\inst{1,2} ,
    B. Berche\inst{2}
    \And P. Butera\inst{3}
    }
\shortauthor{LN Shchur, B. Berche and P. Butera}

\institute{
    \inst{1}  Landau Institute for Theoretical Physics,
    Russian Academy of Sciences, \\
    Chernogolovka 142432, Russia\\
    \inst{2} Laboratoire de Physique des Mat\'eriaux, UMR CNRS 7556,
    Universit\'e Henri Poincar\'e, Nancy 1,\\
    B.P. 239, F-54506  Vand\oe uvre les Nancy Cedex, France\\
    \inst{3} Istituto Nazionale di Fisica Nucleare,
    Sezione di Milano-Bicocca,\\
    Piazza delle Scienze 3, 20126, Milano, Italia\\
    }

\abstract{Monte Carlo (MC) and series expansion (SE) data
    for the energy, specific heat, magnetization and susceptibility of
    the two-dimensional 4-state Potts model in the vicinity of the
    critical point are analysed.
    The role of logarithmic corrections is discussed and an approach is
    proposed in order to account numerically for these corrections
    in the determination of critical amplitudes.
    Accurate estimates of universal amplitude ratios $A_+/A_-$,
    $\Gamma_+/\Gamma_-$, $\Gamma_T/\Gamma_-$ and $R_C^\pm$ are given,
    which arouse new questions with respect to previous works.
}

\begin{document}

\maketitle

%%%%%%%%% INTRODUCTION %%%%%%%%%
{\em Introduction.}
The concept of universality is of fundamental importance in the
theory of phase transitions.
Critical exponents and critical amplitudes describe the
leading singularities
of physical quantities
in the vicinity of the critical point,
\be M_-(\tau) \approx B (-\tau)^\beta,\
\chi_\pm(\tau) \approx \Gamma_\pm \tmod^{-\gamma},\
C_\pm(\tau) \approx \frac{A_\pm}{\alpha}\tmod^{-\alpha},
\label{k-crit}
\ee
($\tau=(T-T_c)/T$ is the reduced temperature
and the labels
$\pm$ refer to the high-temperature (HT) and low-temperature (LT) sides of
the critical temperature $T_c$)
and universal combinations of
critical amplitudes~\cite{PrivmanHohenbergAharony91},
as well as critical exponents
characterize the universality class of the model.
For the Potts models with $q>2$, in
addition to the above mentioned quantities, a transverse
susceptibility is defined
in the LT phase.

Analytical results for the critical amplitudes for the $q$-state
Potts models with $q=1$, $2$, $3$, and $4$ were obtained by Delfino
and Cardy~\cite{DelfinoCardy98}, using the two-dimensional
scattering field theory of Chim and
Zamolodchikov~\cite{ChimZamolodchikov92}. In the case of the 4-state
Potts model, the approach of Ref.~\cite{DelfinoCardy98} leads
to the universal susceptibility amplitude ratios
$\Gamma_+/\Gamma_-=4.013$ and $\Gamma_T/\Gamma_-=0.129$. Monte Carlo
(MC) simulations also reported in~\cite{DelfinoBarkemaCardy00}
did not confirm conclusively these predictions.
Another MC study  due to Caselle et al.~\cite{CaselleTateoVinci99}
leads to  $\Gamma_+/\Gamma_-=3.14(70)$, which is below
the theoretical
prediction of Delfino and Cardy. %, in spite of a fitting procedure with
%non universal logarithmic corrections.
More recently Enting and Guttmann analysed new (longer) series expansions
for $q=3$ and $q=4$ obtained by the finite lattice
method~\cite{EntingGuttmann03}. Their estimates
$\Gamma_+/\Gamma_-=3.5(4)$ and
$\Gamma_T/\Gamma_-=0.11(4)$ for $q=4$
are in slightly better agreement with the results
of~\cite{DelfinoCardy98} and~\cite{DelfinoBarkemaCardy00}.
An analysis by differential approximants however is successful  only for
$q=3$ where the corrections to scaling are represented by pure powers,
but meets with some difficulty in the $q=4$ case, in which logarithmic
corrections are expected. Therefore they had to resort to a slowly
convergent direct analysis of the asymptotic behaviour of the expansion
coefficients.% with respect to their order.

In this letter we present accurate Monte Carlo data
supplemented by a re-analysis of the extended series derived
in~\cite{EntingGuttmann03}.
We are essentially concerned with the universal combinations
\begin{equation}
\frac{A_+}{A_-}, \;\; \frac{\Gamma_+}{\Gamma_-}, \;\;
\frac{\Gamma_T}{\Gamma_-}. \label{univ-rat}
\end{equation}
\revision{To the various critical amplitudes of interest
$A_\pm,\dots$ we have associated appropriately defined   "effective
amplitudes", namely temperature dependent quantities $A_\pm(\tau),
\dots$ which take as limiting values, when $|\tau|\to 0$   the
critical amplitudes  $A_\pm,\dots$. By analogy we have also
considered "effective ratios" of critical amplitudes which tend to
universal ratios as $\tau \rightarrow 0$ and exhibit smoother
behaviours in the vicinity of the critical temperature than the
quantities themselves.} Considering effective ratios would even
eliminate logarithmic corrections from the fit in the case of
4-state Potts model in absence of regular contributions, which
unfortunately do exist! We also use the self-duality relation to
check explicitly the cancellation of the dominant corrections to
scaling in the case of the energy density evaluated at dual
temperatures.

%%%%%%%%% q=4 %%%%%%%%%
{\em Model and observables.}
The Hamiltonian of the Potts model reads as
 $   H = - \sum_{\langle ij \rangle}\delta_{s_i s_j}\; $,
where $s_i$  takes integer values between $0$ and
$q-1$, and the sum is restricted to the nearest-neighbor sites
 $\langle ij \rangle$
 on the  lattice. The partition function is defined by
 $   Z =  \sum_{conf} {\rm e}^{-\beta H}$.
On the square lattice, in zero field, the model is self-dual.
The  duality relation
 $   \left( e^\beta - 1 \right) \left( e^{\beta^*} - 1 \right)=q$
 determines  the critical value of the inverse temperature
$\beta_c=\ln (1+\sqrt{q})$.
Dual reduced temperatures $\tau$ and $\tau^*$ can be defined by
$\beta=\beta_c(1-\tau)$ and $\beta^*=\beta_c(1+\tau^*)$.

In our simulations we use the Wolff algorithm~\cite{Wolff89} for square
lattices of linear sizes $L=20$, $40$, $60$,
$80$, $100$ and $200$ with periodic boundary conditions.
Starting from an ordered state, we let the system equilibrate in
$10^5$ steps measured by the number of flipped clusters. The
averages are computed over $10^6$---$10^7$ steps.
The data are measured in a range of reduced
temperatures called the ``critical window'' and defined as follows:
the lower limit is reached when $\tmod^{-\nu}$ reaches the size $L$
of the system, and the upper limit of the critical window is fixed
for convenience when the corrections to scaling in the Wegner
asymptotic expansion~\cite{Wegner72} do not exceed a few percent,
say $2-3 \%$, of the leading critical behaviour \Eq{k-crit}
(forgetting about the logs). This definition avoids finite size
effects which would
otherwise make our analysis more complex.

The order parameter of a microstate ${\tt M}({\tt t})$ is evaluated
at the time $\tt t$ of the simulation as $    {\tt
M}=\frac{qN_m/N-1}{q-1}$, where $N_m$ is the number of sites $i$
with $s_i=m$ and $m\in [0,...,q-1]$ is the spin value  of the
majority state. $N=L^2$ is the total number of spins. The thermal
average is denoted $M=\<{\tt M}\>$. Thus, the longitudinal
susceptibility in the LT phase is measured by the
fluctuation of the majority spin orientation $k_BT \chi_-=\langle
N_m^2\rangle-\langle N_m\rangle^2$ and the transverse susceptibility
is
defined in the LT phase in terms of the fluctuations
of the
minority of the spins $ k_BT\chi_T=\frac{1}{(q-1)}\sum_{\mu\ne m}
    (\langle N_\mu^2\rangle-\langle N_\mu\rangle^2)$,
while in the HT phase $\chi_+$ is given by the
fluctuations in all $q$ states, $
k_BT\chi_+=\frac{1}{q}\sum_{\mu=0}^{q-1}
    (\langle N_\mu^2\rangle - \langle N_\mu\rangle^2)$,
where $N_\mu$ is the number of sites with the spin in the state $\mu$.
This definition of the
susceptibility is, in both phases, completely consistent with the
available series expansion data~\cite{ShchurButeraBerche02}.
The internal energy density of a microstate is calculated as
$    {\tt E}=-\frac{1}{N} \sum_{\langle ij \rangle}\delta_{s_i s_j}$,
 and its ensemble average is denoted as $E=\<{\tt E}\>$.
 The specific heat measures the energy fluctuations,
$    (k_BT)^{2}C=-\frac{\partial  E }{\partial\beta}
    =\left(\langle {\tt E}^2 \rangle - \langle {\tt E}
\rangle^2 \right)$.

Our MC study of the critical amplitudes will be supplemented by a
 reanalysis of the HT and LT
expansions  recently calculated through remarkably high orders by
Enting, Guttmann and
coworkers~\cite{BriggsEntingGuttmann94,EntingGuttmann03}. In terms
of these series, we can compute the effective critical amplitudes
for the susceptibilities, the specific heat and the magnetization
and extrapolate them by the current resummation techniques, namely
simple Pad\'e approximants (PA)  and differential approximants (DA)
properly biased with the exactly known critical temperatures and
critical exponents. The LT expansion, expressed in terms of the
variable $z= \exp(-\beta)$, extends through  $z^{43}$ in the case of
$E$ , $z^{59}$ in the case of $\chi_-$
($z^{47}$ for $\chi_T$)  and $z^{43}$ for $M$.
The HT expansions, computed in terms of the variable
$v=(1-z)/(1+(q-1)z)$, extend to $v^{43}$ in the case of $E$,
and $v^{24}$ for the $\chi_+$. As a general remark on our
series analysis, we may point out that the accuracy of the amplitude
estimates is questionable, since the mentioned resummation methods
cannot reproduce the expected logarithmic corrections to scaling and
therefore the extrapolations to the critical point are uncertain. In
this case we have also tested a somewhat unconventional use of DA's:
in computing the effective amplitudes, we only retain DA estimates
outside some small vicinity of the critical point, where they appear
to be stable and reliable. Finally we perform the extrapolations by
fitting these data to an asymptotic form which includes logarithmic
corrections.

%%%%%%%%%%%%%%%%%% Logs
{\em Logarithmic corrections.}
In the usual parametrization $\cos(\pi y/2)={\scriptsize\frac 12}\sqrt q$
in terms of which the scaling dimensions are known,
we have $y=0$ at $q=4$ and the second thermal
exponent~\cite{Nienhuis84,DotsenkoFateev84} $\yTdeux=-4y/3(1-y)$ vanishes.
Accordingly, the leading power-behaviour of the magnetization
(and of other physical quantities)
 is modified~\cite{CardyNauenbergScalapino80}
by a logarithmic factor
 \begin{equation}
    M_-(-|\tau|)=B \tmod^{1/12}(-\ln\tmod)^{-1/8}
    {\cal F}_{corr}(-\ln\tmod),
    \label{C4}
\end{equation}
and a correction function ${\cal F}_{corr}(-\ln\tmod)$
contains terms with integer powers of $(-\ln\tmod)$,  $
\ln (-\ln\tmod)/(-\ln\tmod)$,\dots
Non-integer power corrections may also occur due to the higher
(irrelevant) thermal
exponents~\cite{Nienhuis84,DotsenkoFateev84,denNijs79,Pearson80}
$\yTn$ or to other irrelevant fields, but let us first discuss the
form of the logarithmic terms.
Extending  the pioneering
works of Cardy, Nauenberg and
Scalapino (CNS)~\cite{CardyNauenbergScalapino80,NauenbergScalapino80},
Salas and Sokal (SS)~\cite{SalasSokal97}
obtained a slowly
convergent  expansion of ${\cal F}_{corr}(-\ln\tmod)$
 in logs, e.g. for the magnetization:

\begin{equation}
    \label{eq-C_SS}
    M_-(-|\tau|)=B \tmod^{1/12}(-\ln\tmod)^{-1/8}\left[
    1-\frac 3{16}\frac{\ln(-\ln\tmod)}{-\ln\tmod}
    +O\left(\frac 1{\ln\tmod}\right)\right].
\end{equation}

We provide below a re-examination of this and similar quantities.
The non-linear RG equation for the relevant thermal and magnetic
fields $\phi$ and $h$, with corresponding
RG eigenvalues $\yT $ and $\yH $, and
the marginal dilution field $\psi$, are given by
\begin{eqnarray}
    \frac{d\phi }{d \ln b}&=&(\yT +\yTpsi \psi )\phi
        ,\label{ap-eq2}\\
    \frac{dh }{d \ln b}&=&(\yH +\yHpsi \psi)h,
        \label{ap-eq3}\\
    \frac{d\psi }{d \ln b}&=&g(\psi)%=
        \label{ap-eq1}
\end{eqnarray}
where $b$ is the length rescaling factor and $l=\ln b$.
The function $g(\psi)$ may be Taylor expanded,
$g(\psi)=\ypsideux \psi^2(1+\frac{\ypsitrois }{\ypsideux }\psi+\dots)$.
Accounting for marginality of the dilution field, there is no
linear term at $q=4$.
The first term has been considered by Nauenberg and
Scalapino~\cite{NauenbergScalapino80}, and later by
Cardy, Nauenberg and Scalapino~\cite{CardyNauenbergScalapino80}.
The second term was introduced
by Salas and Sokal~\cite{SalasSokal97}.
For convenience, we slightly change the notations of SS,
denoting by $y_{ij}$ the coupling coefficients between the
scaling fields $i$ and $j$.
These parameters
take the values  $\yTpsi =3/(4\pi)$,
$\yHpsi =1/(16\pi)$, $\ypsideux =1/\pi$ and
$\ypsitrois =-1/(2\pi^2)$~\cite{SalasSokal97}, while
the relevant scaling dimensions are
$\yT =3/2$ and $\yH =15/8$.

The fixed point is at $\phi=h=0$. Starting from initial conditions
$\phizero$, $\hzero$, the relevant fields grow exponentially with
$l$. The field $\phi$ is analytically related to the temperature,
so the temperature behaviour follows from the renormalization flow
from $\phizero\sim|\tau|$ up to some $\phil=O(1)$ outside the
critical region. Notice also that the marginal field $\psil$
remains of order $O(\psizero)$ and $\psizero$ is negative,
$|\psizero|=O(1)$. In zero magnetic field, under a change of length
scale, the singular part of the free energy density transforms
according to \be
    f(\psizero,\phizero)=e^{-Dl}f(\psil,\phil),\label{ap-eq4}
\ee where $D=2$ is the space dimension. \revision{Solving
\Eq{ap-eq2} and (\ref{ap-eq1}) leads to
\be
    l=-\frac 1{\yT }\ln x+\frac{\yTpsi }{\yT \ypsideux }
    \ln z,
    \label{ap-eq5}
\ee where $z=\frac{\psizero}{\psil}\frac{\ypsideux +\ypsitrois
\psil}{\ypsideux +\ypsitrois \psizero} $ and $x=\phizero/\phil$} (for
brevity we will denote $\nu=1/\yT=\frac 23$, $\mu=\frac{\yTpsi
}{\yT\ypsideux }=\frac 12$) and we deduce the following behaviour
for the free energy density in zero magnetic field in terms of the
thermal and dilution fields,
\begin{eqnarray}
    f(\phizero,\psizero)&=& x^{D\nu}\ \!
    z^{-D\mu}
    f(\phil,\psil).
    \label{ap-eq6}
\end{eqnarray}
The other thermodynamic properties follow from
derivatives with respect to
the scaling fields, e.g.
$E(\phizero,\psizero)=\frac
\partial{\partial \phizero}f(\psizero,\phizero)
=x^{D\nu-1}\ \! z^{-D\mu}E(\phil,\psil)$. What appears extremely
useful  is that the dependence on the quantity $ z$ cancels (due to
the scaling relations among the critical exponents) in appropriate
effective ratios.  This quantity $z$ is precisely the only one which
contains the log terms in the 4-state Potts model, and thus we may
infer that {\em not only the leading log terms}, but {\em all the
log terms hidden in the dependence on the marginal dilution field}
disappear in the conveniently defined effective ratios. Now we
proceed by iterations of \Eq{ap-eq5} and
 eventually we get for the full correction
 to
scaling variable the {\em heavy} expression $    z=
    {\rm const}\times
    (-\ln|\tau|) \ \!{\cal E}(-\ln|\tau|)
\ \!    {\cal F}(-\ln|\tau|)$, where ${\cal E}(-\ln|\tau|)$
is a universal function

\begin{eqnarray}
\label{eqEfct}
{\cal E}(-\ln|\tau|)&=&
\left(
        1+\frac 34
        \frac{\ln(-\ln|\tau|)}
            {-\ln|\tau|}
    \right)
    \left(1-\frac 34
    \frac{\ln(-\ln|\tau|)}{-\ln|\tau|}
    \right)^{-1}
    \left(
    1+\frac 34\frac{1}{(-\ln|\tau|)}
    \right)
\end{eqnarray}
while ${\cal F}(-\ln|\tau|)$ is a function of the variable
$(-\ln|\tau|)$ only, {\em where
non universality enters through
the constant $\psizero$}. Remember here that $x\simeq|\tau|$.

In a given range of values of the $\tau$, the function ${\cal
F}(-\ln|\tau|)$ should be fixed {\em and the only freedom is to
include background terms and possibly additive corrections to
scaling} coming from irrelevant scaling fields. Among the additive
correction terms, we may have those of the thermal sector
$\Delta_{\phi_n}=-\nu \yTn$, where the RG eigenvalues are
$\yTn=D-\frac 12 n^2, \ n=1,2,3,\dots$~\cite{DotsenkoFateev84}. The
first dimension $\yTun=\yT=3/2$ is the temperature RG eigenvalue.
The next one is $\yTdeux=0$ and this leads to the appearance of the
logarithmic corrections, such that the first Wegner irrelevant
correction to scaling in the thermal sector is
$\Delta_{\phi_3}=-\nu\yTtrois=5/3$. In the magnetic sector, the RG
eigenvalues are given by $\yHn=D-\frac 18 (2n-1)^2$. The first
dimension $\yHun=\yH=15/8$ is the magnetic field RG eigenvalue. The
second one is still relevant, $\yHdeux=7/8$, and it could lead, if
admissible by symmetry, to corrections generically governed by the
{\em difference} of relevant eigenvalues $(\yHun-\yHdeux)/\yT=2/3$.
The next contribution comes from $\yHtrois=-9/8$ and leads to a
Wegner correction-to-scaling exponent
$\Delta_{h_3}=-\nu\yHtrois=3/4$. Eventually, spatial inhomogeneities
of primary fields (higher order derivatives) bring the extra
possibility of integer correction exponents $y_n=-n$ in the
conformal tower of the identity. The first one of these irrelevant
terms corresponds to a Wegner exponent $\Delta_1=-\nu (-1)=2/3$ and
it is always present. We may thus possibly include the following
corrections: $|\tau|^{2/3}$, $|\tau|^{3/4}$, $|\tau|^{4/3}$,
$|\tau|^{5/3}$, \dots, the first and third ones being always
present,
 while the other corrections depend
on the symmetry properties of the observables.

In the Baxter-Wu model, which belongs to the $4-$state Potts model
universality class~\footnote{It was proposed in
Ref.~\cite{NauenbergScalapino80} that $\psi_0=0$ in the Baxter-Wu
model and there are no log-corrections. Later Kinzel et
al.~\cite{Kinzel81} gave supporting considerations.}, the
magnetization obeys the asymptotic form~\cite{Joyce75a,Joyce75b} $
M_-(-\tmod)=B\tmod^{1/12}(1+{\rm const}\times\tmod^{2/3}
    +{\rm const}'\times\tmod^{4/3})$.
Caselle et al.~\cite{CaselleTateoVinci99} also fit the magnetization
with a $|\tau|^{2/3} $ term.

%%%%%%%%%%%%%%%%%%%%%%%%%%
{\em Numerical results.} We eventually deduce the behaviour of the
magnetization
\begin{eqnarray}
    &&M_-(-|\tau|)=B|\tau|^{1/12}(-\ln|\tau|)^{-1/8}
    \left[
    \left(1+\frac 34\frac{\ln(-\ln|\tau|)}{-\ln|\tau|}\right)
        \right.\quad\nonumber\\
    &&\left.
    \ \quad
    \left(1-\frac 34\frac{\ln(-\ln|\tau|)}{-\ln|\tau|}\right)^{-1}
    \left(1+\frac 34\frac{1}{-\ln|\tau|}\right)
    {\cal F}(-\ln|\tau|)
        \right]^{-1/8}(1+a\tmod^{2/3}+\ldots).\label{eq-M_final}
\end{eqnarray}
Note that the whole bracket corresponds to the correction function
of Eq.~(\ref{C4}). It is unsafe (for numerical purposes) to expand
it, since the correction term is not small enough in the accessible
temperature range $|\tau|\simeq 0.05-0.25$. We have thus to extract
an effective function ${\cal F}_{eff}(-\ln|\tau|)$ which mimics the
real one ${\cal F}(-\ln|\tau|)$ in the convenient temperature range.
Defining various effective magnetization amplitudes at different
levels of accuracy, namely
$B^{(1)}_{eff}(-|\tau|)=M_-|\tau|^{-1/12}(-\ln|\tau|)^{1/8}$ with
the CNS leading log term, $B^{(2)}_{eff}(-|\tau|)=
M_-|\tau|^{-1/12}(-\ln|\tau|)^{1/8}\left(1-\frac 3{16}\frac
{\ln(-\ln\tmod)}{-\ln\tmod}\right)^{-1}$ with the SS correction or
$B^{(3)}_{eff}(-|\tau|)=M_-|\tau|^{-1/12}[-\ln|\tau|\ \! {\cal
E}(-\ln \tmod)]^{1/8}$ with our universal corrections, we are unable
to recover a sensible \be
B(1+a\tmod^{2/3}+b\tmod^{4/3})\label{fit1}\ee behaviour. Of course,
it is possible to fit the data to any of these expressions in a
given range of temperatures, but the coefficients $a$ and $b$ thus
obtained strongly depend on the temperature window and this is not
acceptable. Improvement is achieved through the following type of
fit (instead of Eq.~(\ref{fit1})) \be
    B^{(3)}_{eff}(-|\tau|)=B\left(1+\frac{C_1}{-\ln|\tau|}
+\frac{C_2\ln(-\ln|\tau|)}{(-\ln|\tau|)^2}\right)^{1/8}
(1+a\tmod^{2/3}).\label{eq-BeffApp} \ee The function ${\cal
F}(-\ln|\tau|)$ in Eq.~(\ref{eq-M_final}) now takes the approximate
expression
%\be
${\cal F}(-\ln|\tau|)\simeq \left(1+\frac{C_1}{-\ln|\tau|}
+\frac{C_2\ln(-\ln|\tau|)}{(-\ln|\tau|)^2}\right)^{-1}$. What is
remarkable is the stability of the fit to \Eq{eq-BeffApp}. Analysing
MC data, we obtain (fit a) $C_1=-0.757(1)$ and $C_2=-0.522(11)$
which yields an amplitude $B=1.1570(1)$. It is also possible to try
a simpler choice in the narrow temperature window, fixing $C_2=0$
and approximating the whole series by the $C_1-$term only (now
$C_1=-0.88(5)$, called fit b), which then leads to a very close
magnetization amplitude $B=1.1559(12)$. An analysis of SE data gives
very similar results. By the way, in the case of the magnetization
the coefficient $b$ is found to be almost zero and we did not
include it in \Eq{eq-BeffApp}~\cite{CSP07}. Note that these
estimates follow from a coherent analysis of both MC data and SE
extrapolations.
\revision{
 The errors reported are the standard deviations resulting
from the fits, since our definition of the temperature window is
such that there is no finite-size-effect in the  $\tau-$range
considered.
For MC data we perform weighted fits (i.e. each point is
weighted with the inverse statistical error of the point) while the
fits of the  SE data are unweighted. Our major improvement
(compared to previous references using MC and/or SE data) is not in
the quality of the data themselves, but in the functional form of
fit employed which incorporates in an effective function
(dependent on the temperature window)
 the effect of the {\em
non-universal} part of the series of log terms, all universal terms
being explicitly taken into account. Nevertheless,
another source of error  comes from the
effective function itself. We can estimate this additional
error in the following way: we compute the amplitudes when
changing the coefficients $C_1$ and $C_2$
of an amount as large as half to twice their
optimal values reported above,
leading to two estimates for the amplitudes. We
arbitrarily define the difference of these estimates
  as the additional error.
$\Gamma_-$ is the only amplitude for which this error is
significant.}

We thus obtain a closed
expression for the dominant logarithmic corrections which
is more suitable than previously proposed forms
to describe the temperature range accessible in a numerical study:
\bey
    {\rm Obs.}(\pm|\tau|)
        &\simeq&{\rm Ampl.}\times
    \tmod^{\triangleleft}\times
    [{\cal E}(-\ln\tmod){\cal F}(-\ln\tmod)]^{\#}\nonumber\\
    & &\qquad\times(1+{\rm Corr.\ terms})+
    %&\ &\qquad\qquad+
    {\rm\  Backgr.\ terms},
        \label{eq-Obs_us}\\
    {\rm Corr.\ terms}&=&a \tmod^{2/3}+b \tmod^{4/3}+\dots %b_\pm\tmod+\dots
    ,\label{eq-Obs2}\\
    {\rm Backgr.\ terms}&=&D_0+D_1\tmod+\dots
\label{eq-Obs} \eey where ${\triangleleft}$ and $\#$ are exponents
which depend on the observable considered, and take the values
$1/12$ and $-1/8$, respectively, in the case of the magnetization.
The dots represent higher order terms which theoretically do exist,
but practically
do not need to be included.% in the fits.

\begin{figure}
\centerline{\epsfig{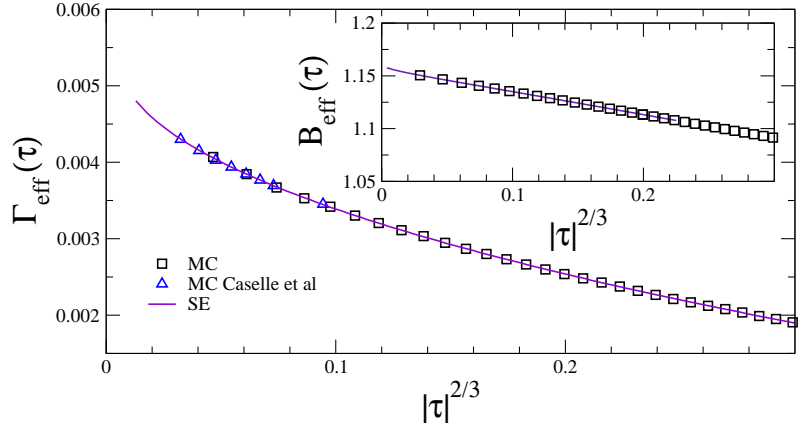}}
\caption{\revision{
Effective susceptibility amplitude (and magnetization amplitude
in the inset) in the LT phase.}}
\label{f.lbl}
\end{figure}

The susceptibility (see Fig.~\ref{f.lbl}) and the energy density can
also be fitted to the expression above. Our results are summarized
in table~\ref{tab-q4-res}. The efficiency of  the fits relies on the
asymptotic form \Eq{eq-Obs_us} which in our opinion is based on
sufficiently safe theoretical grounds. Its validity can furthermore
easily be checked (indirectly) through the computation of effective
amplitude ratios for which all logarithmic corrections have to
cancel. A specific example is given by the leading behaviour of the
energy density ratio. The values $E(\beta)$ and $E(\beta^*)$ of the
internal energy at dual temperatures are related through $
\left(1-e^{-\beta}\right) E(\beta) + \left(1-e^{-\beta^*}\right)
    E(\beta^*)=-2$.
Defining the quantity
$A_+/A_-=(E(\beta)-E_0)\tau^{\alpha-1}/(E_0-E(\beta^*))(\tau^*)^{\alpha-1}$,
the constant $E_0$ being the value of the energy at the transition
temperature~\cite{Baxter}, $E_0=E(\beta_c)=-1-1/\sqrt{q}$, we may
expand close to the transition point
$A_+(\tau)/A_-(\tau*)=1+(3-\alpha)\alpha_q \tau+O(\tau^{1+\alpha})$
with $\alpha_q=
-E_0\beta_ce^{-\beta_c}=\frac{\ln(1+\sqrt{q})}{\sqrt{q}}$. This
relation, checked numerically, shows that the leading corrections to
scaling vanish.

The universal combinations of amplitudes follow from the results
listed in table~\ref{tab-q4-res} and are summarized in
table~\ref{tab-3-resbisq4}. Fits a and b in these tables refer to
the two possible choices for the constants $C_1$ and $C_2$ in ${\cal
F}(-\ln\tmod)$ as explained above.

\begin{table}[h]
\caption{Critical amplitudes in the 4-state
    Potts model. The amplitudes reported correspond to %an average between
the estimates which follow from the analysis of MC data and of SE
data with both types of fits.  \revision{The
second figures in
parenthesis for $\Gamma_-$ refer to the additional error discussed in the
text.}}
\center\scriptsize%\footnotesize
\begin{tabular}{lllllllll}
\hline\noalign{\vskip-1pt}\hline\noalign{\vskip2pt}
 F fit \# & $B$ & $\Gamma_+$ & $\Gamma_-$ & $\Gamma_T$   \\
\noalign{\vskip2pt}\hline
MC a &\revision {1.1570(1)}  &\revision {0.03144(15)} &\revision {0.00454(2)(20)}  &\revision {0.00076(1)}\\
MC b &\revision {1.1559(12)} &\revision {0.03178(30)} &\revision {0.00484(3)(5)}   &\revision {0.00073(1)}\\
SE a &\revision {1.1575(1)}  &\revision {0.03041(1)}  &\revision {0.00483(1)(20)}  &\revision {0.00073(1)}\\
SE b &\revision {1.1575(1)}  &\revision {0.03039(1)}  &\revision {0.00493(1)(5)}   &\revision {0.00073(1)}\\
\hline\noalign{\vskip-1pt}\hline
\end{tabular}
\label{tab-q4-res}
\end{table}

\begin{table}[h]
\caption{Universal combinations of the critical amplitudes in the 4-state
    Potts model. \revision{In the last two rows,
the figure on the left follow from MC results and that on the right
    from SE results.}}
\center\scriptsize%\footnotesize
\begin{tabular}{lccc}
\hline\noalign{\vskip-1pt}\hline\noalign{\vskip2pt}
$A_+/A_-$ & $\Gamma_+/\Gamma_-$ & $\Gamma_T/\Gamma_-$ %& $R_C^+$ & $R_C^-$
& source
\\ \noalign{\vskip2pt}\hline
   $1.$       &  $4.013$            & $0.129$ &
     \cite{DelfinoCardy98,DelfinoBarkemaCardy00}\\
  $-$        &  $3.14(70)$              &       $-$ &
     \cite{CaselleTateoVinci99}\\
   $-$       &  $3.5(4)$   & $0.11(4)$ &
     \cite{EntingGuttmann03}\\
$1.00(1)$   &  \revision {$ 6.93(6)(35) - 6.30(2)(27)  $}         &
\revision {$0.167(3)(9) - 0.151(2)(8)  $ }     &
     {fit a}\\
$1.00(1)$   &  \revision {$ 6.57(10)(13) - 6.16(1)(6) $}         &
\revision {$ 0.151(3)(4) - 0.148(2)(4)  $} &
     {fit b}\\
\hline\noalign{\vskip-1pt}\hline\noalign{\vskip2pt}
\end{tabular}
\label{tab-3-resbisq4}
\end{table}

%%%%%%%%%%%%%%%%%%%%%%%%%%
{\em Conclusion.} The main outcome of this work is the surprisingly
high values of the ratios $\Gamma_+/\Gamma_-$, $\Gamma_T/\Gamma_-$
and $R_C^+$, clearly far above the predictions of Delfino and Cardy.
\revision{Note that our results are also supported by a direct
extrapolation of {\em effective amplitude ratios} for which most of
the corrections to scaling disappear. In the case of the conflicting
quantities, this technique leads to $\Gamma_+/\Gamma_-=6.6(3)$ and
$6.5(1)$, and $\Gamma_T/\Gamma_-=0.160(8)$ and $0.152(2)$, using
respectively fits a and b to fit the MC data. The corresponding
figures resulting from fits of SE data are
$\Gamma_+/\Gamma_-=6.30(1)$ and $6.16(1)$, and
$\Gamma_T/\Gamma_-=0.151(3)$ and $0.148(3)$. Note that the
additional source of error is not taken into account in these
estimates.}

We believe that our fitting procedure is reliable, and since the
disagreement with theoretical calculations can hardly be resolved,
we suspect that the discrepancy might be attributed to the
assumptions made in Ref.~\cite{DelfinoCardy98} in order to predict
the susceptibility ratios. Even more puzzling is the fact that
Delfino and Cardy argue in favour of a higher robustness of their
results for $\Gamma_T/\Gamma_-$ than for $\Gamma_+/\Gamma_-$, but
the disagreement is indisputable in both cases.

Finally, in favour of our results, one may mention a work of W. Janke
and one of
us (LNS) on the amplitude ratios in the Baxter-Wu model (in the
4-state Potts model universality class), according to which
$\Gamma_+/\Gamma_-\simeq 6.9$~\cite{BBJS06}.
\revision{These results, obtained from an analysis of
MC data} show a similar discrepancy with Delfino and Cardy's results
and a further analysis still seems to be necessary.

\stars We gratefully acknowledge discussions with A. Zamolodchikov
and W. Janke, correspondence with V. Plechko,  J. Salas and J.L. Cardy, and
we are especially thankful to Malte Henkel for discussions on many aspects
of conformal invariance. We thank the Twinning programme between the
CNRS and the Landau Institute which made possible this cooperation.

%%%%%%%%%%%%%%%%%%%%%%%%%%%%%%%%%%%%%%%%%%%%%%%%%%%%%%%%%%%
%        REFERENCES
%%%%%%%%%%%%%%%%%%%%%%%%%%%%%%%%%%%%%%%%%%%%%%%%%%%%%%%%%%

\end{document}